\documentclass{article}

\usepackage{arxiv}
\usepackage{listings}
\usepackage{xcolor}
\usepackage[utf8]{inputenc} 
\usepackage[T1]{fontenc}    
\usepackage{hyperref}       
\usepackage{url}            
\usepackage{booktabs}       
\usepackage{amsfonts}       
\usepackage{nicefrac}       
\usepackage{microtype}      
\usepackage{lipsum}
\usepackage{graphicx}
\usepackage{multirow}
\graphicspath{ {./images/} }
\usepackage{tikz}
\usepackage{pgfplots}
\pgfplotsset{width=10cm,compat=1.17}

\lstdefinestyle{DOS}{
    backgroundcolor=\color{black},
    basicstyle=\scriptsize\color{white}\ttfamily
}

\definecolor{mGreen}{rgb}{0,0.6,0.6}
\definecolor{mGray}{rgb}{0.7,0.7,0.7}
\definecolor{mPurple}{rgb}{0.58,0,0.82}
\definecolor{backgroundColour}{rgb}{1,1,1}

\lstdefinestyle{CStyle}{
    backgroundcolor=\color{gray!20},   
    commentstyle=\color{mGreen},
    keywordstyle=\color{red},
    numberstyle=\tiny\color{mGray},
    stringstyle=\color{mPurple},
    basicstyle=\footnotesize,
    breakatwhitespace=false,         
    breaklines=true,                 
    captionpos=b,                    
    keepspaces=true,                 
    numbers=left,                    
    numbersep=5pt,                  
    showspaces=false,                
    showstringspaces=false,
    showtabs=false,                  
    tabsize=2,
    language=C
}

\title{Flex Net Sim: A Lightly Manual}

\author{
 Felipe Falcón \\
  Faculty of Engineering and Sciences\\
  Universidad Adolfo Ibáñez\\
  \texttt{ffalcon@alumnos.uai.cl} \\
   \And
 Gonzalo España \\
  Faculty of Engineering and Sciences\\
  Universidad Adolfo Ibáñez\\
  \texttt{gespana@alumnos.uai.cl} \\
  \And
 Danilo Bórquez-Paredes \\
  Faculty of Engineering and Sciences\\
  Universidad Adolfo Ibáñez\\
  \texttt{danilo.borquez.p@uai.cl} \\
}

\begin{document}
\maketitle
\begin{abstract}
A common problem in elastic optical networks is to study the behavior of different resources allocation algorithms, such as signal modulation formats or quality of service, in optical networks in dynamic scenarios where connections are assigned and released following different traffic profiles. To achieve this, one of the busiest tools is simulators. Normally each research group has its own simulator created entirely by them, which works on a particular simulation scenario, generating multiple versions of the same simulator. For this reason, this project aims to create a tool that allows focusing on the creation of algorithms, generating a common platform for simulation.

We present a C ++ library that contains the most common modules belonging to an event-oriented simulator for flexible grid optical networks. This library allows researchers to worry about algorithm generation rather than maintaining/modifying a simulator. The final product is a library capable of being included in any program written in C ++, allowing the design of resource allocation algorithms through macros used in the same source file of the user that uses the library.
\end{abstract}


\section{Motivation}

The need for communications through the internet is being increasingly demanded~\cite{Al-Tarawneh2021}. In this context, today, it is common to speak of high-speed communications on computers and mobile devices. To make this possible, the infrastructure that supports these transmission rates must have the capacity demanded, which is naturally delivered by networks composed of fiber optics, known as optical networks~\cite{Tzanakaki2020}.

Optical networks comprise a set of nodes connected by another set of optical links. Optical links can carry information within a portion of the spectrum of light on the C-band~\cite{Ferrari2020}. They established communication between a pair of nodes using a specific wavelength within the C-band, which cannot be used at the same time by another pair of nodes on the same link. This last feature is called Wavelength Division Multiplexing (WDM)~\cite{Mukherjee2006}, where each wavelength is a communication channel. When a link does not connect a pair of nodes, they can communicate using a set of links that is known as a route. In addition, the communication established between each pair of nodes uses a specific modulation format~\cite{Lopez2016}. 

In the past, the bandwidth used by each connection in this type of network was fixed, typically 50 GHz~\cite{Yoo2012}. This caused communications that demand little bandwidth to be sub-occupying the spectrum, wasting a big part of the spectrum. This is how, for example, a 10 Gbps communication that is modulated with a BPSK format and used 12.5 GHz, does not occupy all the assigned spectrum, leaving 37.5 GHz free (wasting)~\cite{calderon2020}. 

It is in this context that elastic optical networks (EONs) appeared, which have a flexible spectrum of allocation~\cite{Dixit2021}. Thus, the spectrum available on each link is divided into fundamental units typically called Frequency slot units (FSU). In this context, each connection comprises one or more FSUs, which together satisfy the bandwidth demand of a connection fulfilling a couple of restrictions: continuity and continuity. The continuity constraint states that if a connection uses over one link in the network, then the FSUs used must be the same throughout the entire route. The contiguity constraint instead forces the FSUs used by a connection to be side by side, with no gaps in between.

A common problem to solve is that of assigning a route, a modulation format, and a set of slots to specific connections. We know this problem as Routing, Modulation Level, and spectrum assignment problem (RMLSA)~\cite{Chatterjee2020}. The scenarios in which it solved this problem can be two, depending on the traffic profile used: static and dynamic~\cite{Chadha2019}. In the static scenario, the set of connections to be assigned is known a priori, so it is usually solved with integer linear programming techniques or meta-heuristics . In a dynamic scenario, the connection requests are random, as is the connection time. This scenario is normally solved through implementing heuristics, using an optical network simulator as a simulation tool.

The amount of simulation software related to elastic optical networks is not very large. Among the simulation tools available are EONS~\cite{EONS}, ElasticO++~\cite{elastico}, FlexGridSim~\cite{flexgridsim} and CEONS~\cite{Aibin2015}. EONS is a dynamic traffic simulator written in java, with algorithms previously designed for the RSA problem. It does not work by incorporating modulation formats into the algorithms and blocking probability as the final metric. It is not a flexible tool for studying new algorithms and new metrics. ElasticO++ is a framework for the simulation of flexible grid optical networks, but it works as an extension of another simulator called OMNET++~\cite{Varga2010}. It assumes a structure in the resource allocation algorithms, so each new algorithm must use that structure. It works with terms such as inheritance and interfaces, making it an extra barrier to entry when using this simulator. Finally, the repository in bitbucket is no longer accessible~\footnote{\url{https://bitbucket.org/Stange/elastico/ }}. FlexGridSim is an extremely simple simulator written in Java, based on WDMSim~\cite{wdmsim}. It has relatively poor documentation, and it does not show how the new algorithms are created, so it is not an easy-to-use tool. CEONS is written in Java with a friendly graphical interface. Its major problem is that it comes with the algorithms already set, so it is not made mainly to create them but to see how they behave under different parameters. ONS is written in Java, and it tries to provide a simple way to implement new RWA / RSA / RMLSA algorithms. It is a complete simulator, but it comes with pre-defined statistics. The programmer must know concepts of inheritance or interfaces, typical of object orientation that can be an entry barrier for using the software. Finally, SNetS It is also written in Java, but it works as a local server that runs the simulation. The algorithms are also pre-programmed, so it does not seem to help to create new ones.

In this paper we present an event-oriented simulation library for elastic optical networks written in C ++. The architecture of the library gives researchers great flexibility to create resource allocation algorithms, as well as to insert variables intended to measure any aspect of the algorithm. Its C ++ programming allows you to take advantage of the performance offered by this language in terms of runtime. 

In Section 2 we presented how the library should be installed. Section 3 shows the major commands that can be used, as well as the general structure of algorithm programming. Section 4 presents examples of algorithms performed using the library, and finally in section 5 we present the conclusions.

\section{Installation and compilation}

We detailed the installation method for the most common operating systems below. There is also a big header file version that can be downloaded directly from the repository\footnote{\url{https://gitlab.com/DaniloBorquez/flex-net-sim}}$^{,}$\footnote{\url{https://daniloborquez.gitlab.io/flex-net-sim/simulator.hpp}}. This type of file usually contains all the code of a system in a single document which can be included within a C / C ++ program. The big header file of the Flex Net Sim library contains all the code necessary for the operation of the library, and you only need to add it to the project in which you want to use it.

\subsection{Dependencies}

The simulation library is written in C ++, using the C ++ 11 standard. For this reason, an obvious dependency is a C ++ compiler installed on the machine where the library will be used. 

The second dependency required is the software build, test, and package tool called CMake~\cite{Cmake}, in its 3.15 version. This tool is used to compile the library using the compiler installed in the operating system, as well as the unit tests that allow to corroborate its operation. 

Finally, the simulation library uses Catch2~\cite{Catch2} for implementing the unit tests. Although this is clearly a dependency, Flex Net Sim uses Cath2’s big header file for testing, which is already embedded within Flex Net Sim. The prior installation of this library on the computer where the simulations will be made is unnecessary.

\subsection{Compiling and installing library}

To compile and install Flex Net Sim we use CMake, following the coming four steps: first we create a folder where we will set up the library, later we prepare the compilation files. Then we compile through the \textit{build} option that CMake has, and finally we install the library using the \textit{install} option of CMake. 

Although the commands to install the library are very similar to each other, we prefer to display them for each selected operating system separately to avoid possible confusion.

\subsubsection{Windows}

The default folder used to install programs on Windows operating systems is Program Files. Cmake installs Flex Net Sim by default in this folder as well. To do this, you must open a windows console as administrator and execute the following commands:
    
    \begin{lstlisting}[] 
    $ mkdir build
    $ cd build
    $ cmake .. 
    $ cmake --build . --config Release
    $ cmake --install .
    \end{lstlisting}

\subsubsection{Unix systems}

On Unix systems, the default folder chosen to install the library is /usr/local/lib/. The following commands written in a terminal allow the installation of the library:

    \begin{lstlisting}[] 
    $ mkdir build
    $ cd build
    $ cmake ..
    $ cmake --build .
    $ sudo cmake --install .
    \end{lstlisting}


\subsection{Compiling and linking to your program}

Using the library in a project written in C ++ is through the \textit{include} keyword, like any other library. If you use the big header file, then just add the library with the double quotes, depending on the location you chose. If you use the previously installed version, then by default it is installed inside a folder called fnsim, so when including it in your project you must refer to this library as shown below.

    \begin{lstlisting}[style=CStyle]
    #include <fnsim/simulator.hpp>\end{lstlisting}
    
To compile the project, we simply called the compiler from the console, taking care that there is a version higher than the C ++ 11 standard installed. One way to ensure this requirement is by adding the corresponding flag to the compilation. If we compile the project using the installed library, then we need to add the -lfnsim flag to the compilation line.

    \begin{lstlisting}[] 
    $ g++ -std=c++11 <your cpp files> -lfnsim
    \end{lstlisting}

With windows, the compilation is a little different. Windows does not have a well-defined folder to store the libraries and includes, depending heavily on the compilers that are installed in the operating system. Therefore, the corresponding flags must be added to the compilation line. The following command assumes a library installation in the path C:/Program Files (x86)/flexible-networks-simulator/. 

    \begin{lstlisting}[] 
     $ g++ -std=c++11 <your cpp files> 
     -I"C:/Program Files (x86)/flexible-networks-simulator/include/" 
     -L"C:/Program Files (x86)/flexible-networks-simulator/lib/" -lfnsim
    \end{lstlisting}

\section{Using library}

In this section we show how to use the Flex Net Sim library in a C++ code.

\subsection{Principal commands}

In order to use Flex Net Sim library to create Flexible Grid Optical Network simulations in your program, we require two elements: a Simulator object and an Allocation function, which corresponds to the desired allocation algorithm.

We must instantiate the Simulator object inside the main function of the program. We can achieve this through three different constructors.
\begin{itemize}
    \item The empty/void constructor, which receives no argument.
    
    \begin{lstlisting}[style=CStyle]
    Simulator sim = Simulator();\end{lstlisting}
    
    \item The Network routes and paths constructor, which receives as first and second argument the path to the JSON files containing information about the desired Network routes and paths, respectively.
    
    \begin{lstlisting}[style=CStyle]
    Simulator sim = Simulator(std::string("routes.json"),
    std::string("paths.json"));\end{lstlisting}
    
    \item The Network routes, paths and bit rate constructor, which receives the same two arguments as the former constructor and the path to a JSON file containing information about the desired bit rates to use as third argument.
    
    \begin{lstlisting}[style=CStyle]
    Simulator sim = Simulator(std::string("routes.json"),
    std::string("paths.json"),std::string("bit_rates.json"));\end{lstlisting}
    
\end{itemize}

All the constructors will instantiate a Simulator object with default values for several attributes (arrive and departure ratios, goal connections, arrival seed, departure seed, bit rate seed, bitrates, among others).

We must define the allocation algorithm in the global scope, between two function macros, BEGIN\_ALLOC\_FUNCTION(<name>) and END\_ALLOC\_FUNCTION, and usually after the includes and before the main function. The name argument in the macro is the desired name for the allocation function to create.

The allocation algorithm will be called every time there is a connection request between any pair of nodes. Within it, we can use various macros, which are listed below:

\begin{itemize}
    \item NUMBER\_OF\_ROUTES returns the number of routes between the source and destination nodes of the current request.
    \item NUMBER\_OF\_LINKS(route) returns the number of links between the source and destination nodes of the current request in a given route. The route is an integer representing the route position, starting from 0.
    \item LINK\_IN\_ROUTE(route,link) returns the reference (pointer) to a given link in a given route between the source and destination nodes of the current request. The route is an integer representing the route position, starting from 0. The link is an integer representing the link position in the specific route, starting from 0.
    \item LINK\_IN\_ROUTE\_SRC(route,link) returns the Id of the source node from the reference to a given link in a given route. The route is an integer representing the route position, starting from 0. The link is an integer representing the link position in the specific route, starting from 0.
    \item LINK\_IN\_ROUTE\_DST(route,link) returns the Id of the destination node from the reference to a given link in a given route. The route is an integer representing the route position, starting from 0. The link is an integer representing the link position in the specific route, starting from 0.
    \item LINK\_IN\_ROUTE\_ID(route,link) returns the Id from the reference to a given link in a given route. The route is an integer representing the route position, starting from 0. The link is an integer representing the link position in the specific route, starting from 0.
    \item REQ\_SLOTS(position) returns the number of required slots of a Bit Rate object inside the Bit Rate vector at a desired position. The position is an integer representing the bitrate position, starting from 0. 
    \item REQ\_BITRATE returns the value (as float) of the Bit Rate (in bits/s) from the Bit Rate object.
    \item REQ\_BITRATE\_STR returns the value (parsed as string) of the Bit Rate (in bits/s) from the Bit Rate object.
    \item REQ\_REACH(position) returns the reach of the required Bit Rate from its reach vector at a desired position. The position is an integer representing the bitrate position, starting from 0.
    \item REQ\_MODULATION(position) returns the modulation format of the required bitrate from its modulation vector at a desired position. The position is an integer representing the bitrate position, starting from 0.
    \item ALLOC\_SLOTS(link,from,to) it is used to Allocate a desired number of slots (starting at "from" position, ending at "to - 1" position) inside a given link whenever appropriate depending on the algorithm to be coded. The link is an integer representing the link id, starting from 0.
    \item SRC returns the Id of the source node in the current request.
    \item DST returns the Id of the destination node in the current request.
    
\end{itemize}

Note that these functions are useful for iterate over all links of all routes and allocating slots in the desired and suitable links. For instance, a loop over NUMBER\_OF\_ROUTES will iterate over all routes. Inside this loop, another loop iterating over NUMBER\_OF\_LINKS(route) can iterate over all the links inside each specific route. This allows any link in any route to be reached with LINK\_IN\_ROUTE(route,link) and thus we can allocate their slots with ALLOC\_SLOTS(link,from,to).

Be aware that it’s possible to call any of the Link’s methods at the end of the LINK\_IN\_ROUTE(route,link) function macro. For instance, if we need the number of slots in a link from a route, it is possible to call LINK\_IN\_ROUTE(route, link)->getSlots(). Also we can add some statistical variables created by us to store any relevant information that we want to save. 

Finally, when it is possible to allocate the required slots or not, depending on the algorithm to be coded, we shall return an ALLOCATED or NOT\_ALLOCATED . These return values are required, as they will be used for calculating the Blocking probability shown on the console log.

For instance, the skeleton of an Allocation Algorithm definition should follow a pattern as shown below:

\begin{lstlisting}[style=CStyle]
    BEGIN_ALLOC_FUNCTION(exampleName)
    {
        for(iterate/traverse routes/links/...)
        {
            ...
            some statistic variable
            ...
            if(condition is met)
            return ALLOCATED;
            ...
        }
        ...
        return NOT_ALLOCATED;
    }
    END_ALLOC_FUNCTION\end{lstlisting}

Once defined the Allocation function and instantiated a Simulator object, then it is required to set the allocation algorithm to be used by the simulation with the function macro USE\_ALLOC\_FUNCTION(<name>, <simulator>). The name argument is the same name used when the desired allocation function was defined and the simulator argument is a simulator object.

For instance, the code below will link the sim Simulator object to the exampleName allocation algorithm in order to use that algorithm to run the simulations afterwards.

\begin{lstlisting}[style=CStyle]
    USE_ALLOC_FUNCTION(exampleName, sim);\end{lstlisting}

So far any default simulator object has the default attributes. If any attribute needs to be changed, we could change it by using any of the attribute setters. 

Once everything is set up correctly for the simulation, we initialize the simulator and we execute it. The simulator not accepts new changes after the initialization through the init() method.

The simulator initialization will properly set the simulation clock to 0 and other needed variables. Once the simulator is run, it will output information corresponding to the principal parameters, and the used algorithm for the simulation. As simulations go on, the Simulator will output the progress of the simulation on the console.

\section{Examples}

The simulator library  was tested with three different algorithms, that were used in~\cite{borquez2018does}. A brief description about the algorithms are showed next:

\begin{itemize}
    \item First Fit (FF): The connection request is allocated in the first spectrum space that match with the required bandwidth. Usually the algorithm goes from the lowest frequency in the spectrum to the highest, and if it found sufficient space that satisfies  requirement, then allocate the resources,
    \item Exact Fit (EF): This algorithm focus on allocate the connection demand in a space in the spectrum that exactly match with the spectrum requirement. When not space that match exactly is found, then it works as a FF algorithm.
    \item First-Last-Fit (FLF): The connection demands are divided into two groups according to a threshold, and they use a First Fit algorithm in a forward or a reverse way. In this work we use the bitrate as threshold, so demands that are less than 100 Gbps use the FF algorithm from the lowest to the highest spectrum slots, meanwhile the connections with a bitrate greater or equal to 100 Gbps use the FF algorithm from highest to slower spectrum slot.    
\end{itemize}

The parameters of the network used are presented in the Table~\ref{tab:sim_params}:

\begin{table}[!h]
\centering
        \begin{tabular}{|c|c|}
            \hline
                \textbf{Parameter}      & \textbf{Value}  
                 \\ \hline
                 Topology       &NSFNet (14 nodes, 42 links)\\ \hline
                Capacity per link           & 320 slots         \\ \hline
                Bitrates               & 10, 40, 100, 400, 1000 Gbps    \\ \hline
                \begin{tabular}[c]{@{}c@{}}Modulation \\ formats\end{tabular}       & \begin{tabular}[c]{@{}c@{}}BPSK, QPSK, 8-QAM,\\ 16-QAM, 32-QAM, 64-QAM\end{tabular}             \\ \hline
        \end{tabular}
\caption{Network Characteristics}
\label{tab:sim_params}
\end{table}

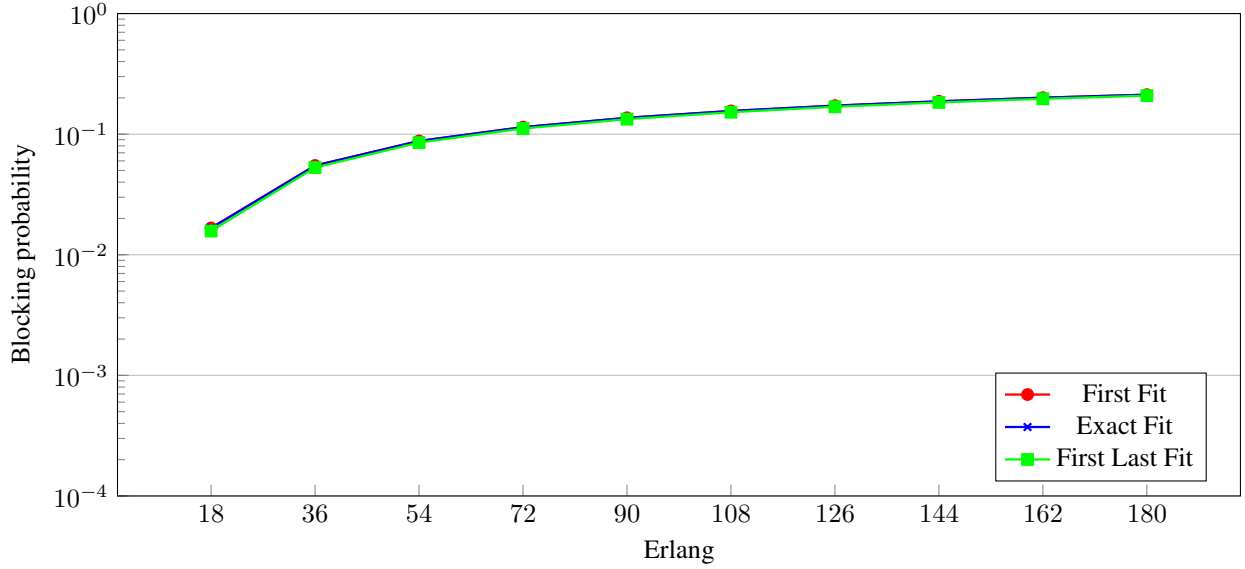
\begin{figure}
    \centering
    \begin{tikzpicture}
    	\begin{axis}[
        	stack plots=false,
        	width=\textwidth,
        	height=8cm,
            ymode=log,
         	ymin=1e-4, ymax=1,
         	legend pos=south east,
         	xtick=data,
        	ylabel = {Blocking probability},
        	xlabel = {Erlang},
        	ymajorgrids=true,
        	xtick pos=left,
            ytick pos=left
        	]
            \addplot[color=red,mark=*, mark options={fill=red},line width=0.3mm] table{alg1_FF.dat};
            \addplot[color=blue,mark=x,line width=0.3mm] table{alg1_EF.dat};
            \addplot[color=green,mark=square*, mark options={fill=green},line width=0.3mm] table{alg1_FLF.dat};
            \legend{First Fit, Exact Fit, First Last Fit}
        \end{axis}
    \end{tikzpicture}
    \caption{Simulations performed with 1 route, all conections with BPSK modulation format and not considering optical reach}
    \label{fig:simple}
\end{figure}

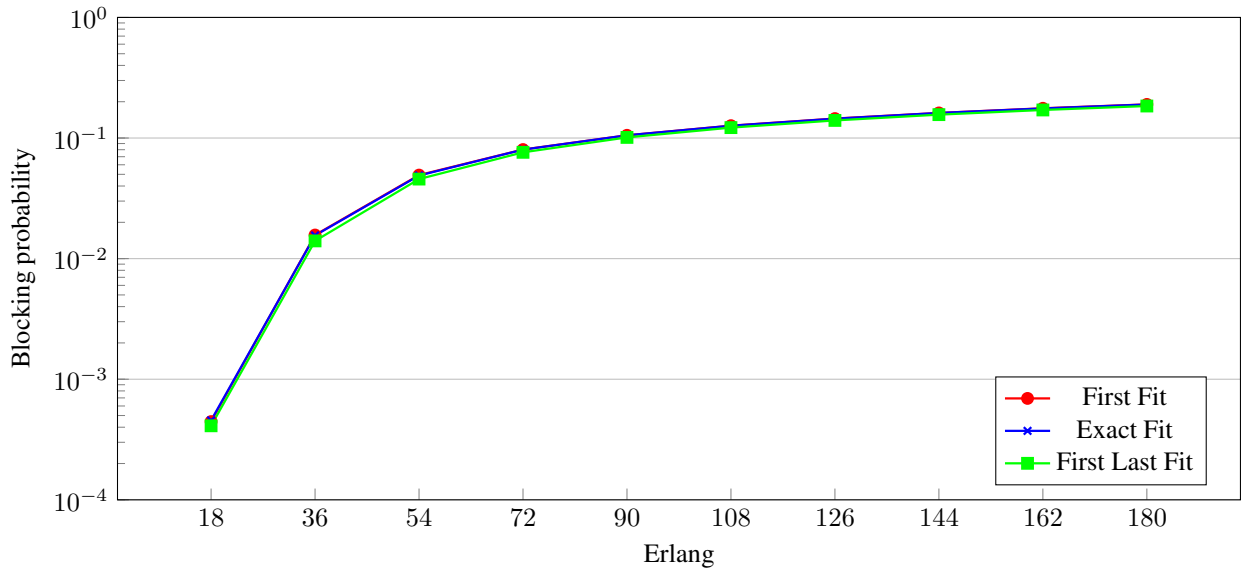
\begin{figure}
    \centering
    \begin{tikzpicture}
    	\begin{axis}[
        	stack plots=false,
        	width=\textwidth,
        	height=8cm,
            ymode=log,
         	ymin=1e-4, ymax=1,
         	legend pos=south east,
         	xtick=data,
        	ylabel = {Blocking probability},
        	xlabel = {Erlang},
        	ymajorgrids=true,
        	xtick pos=left,
            ytick pos=left
        	]
            \addplot[color=red,mark=*, mark options={fill=red},line width=0.3mm] table{alg2_FF.dat};
            \addplot[color=blue,mark=x,line width=0.3mm] table{alg2_EF.dat};;
            \addplot[color=green,mark=square*, mark options={fill=green},line width=0.3mm] table{alg2_FLF.dat};;
            \legend{First Fit, Exact Fit, First Last Fit}
        \end{axis}
    \end{tikzpicture}
    \caption{Simulations performed with 3 routes, all conections with BPSK modulation format and not considering optical reach}
    \label{fig:routes}
\end{figure}

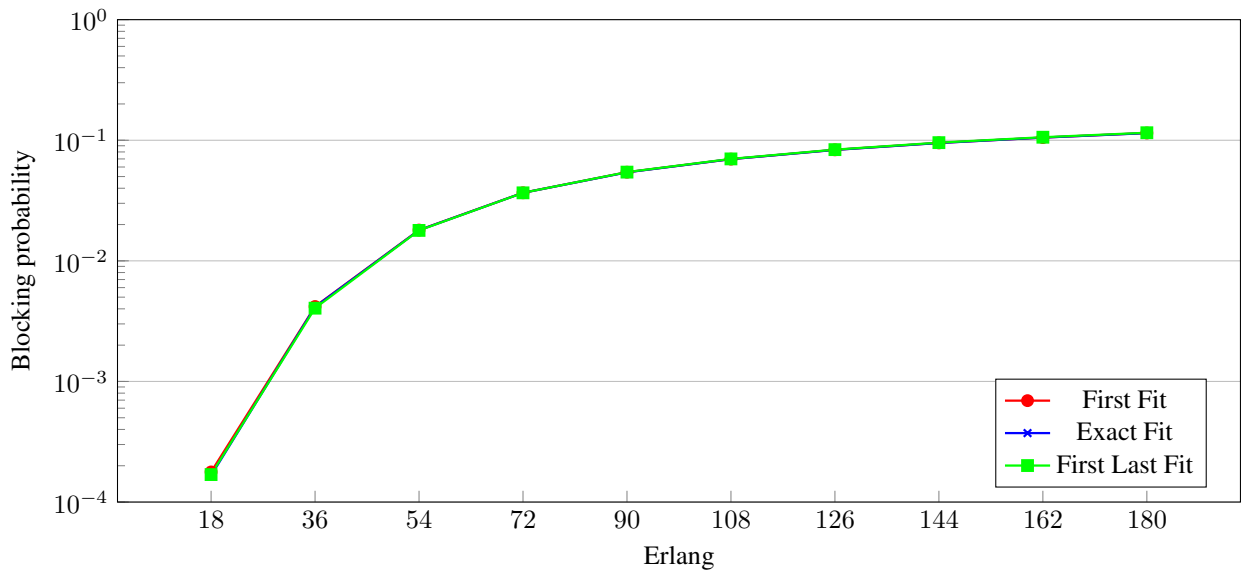
\begin{figure}
    \centering
    \begin{tikzpicture}
    	\begin{axis}[
        	stack plots=false,
        	width=\textwidth,
        	height=8cm,
            ymode=log,
         	ymin=1e-4, ymax=1,
         	legend pos=south east,
         	xtick=data,
        	ylabel = {Blocking probability},
        	xlabel = {Erlang},
        	ymajorgrids=true,
        	xtick pos=left,
            ytick pos=left
        	]
            \addplot[color=red,mark=*, mark options={fill=red},line width=0.3mm] table{alg3_FF.dat};
            \addplot[color=blue,mark=x,line width=0.3mm] table{alg3_EF.dat};;
            \addplot[color=green,mark=square*, mark options={fill=green},line width=0.3mm] table{alg3_FLF.dat};;
            \centering
            \legend{First Fit, Exact Fit, First Last Fit}
        \end{axis}
    \end{tikzpicture}
    \caption{Simulations performed with 3 routes, using modulation formats and optical reachs presented in Table~\ref{tabla:bitrate_modulation_fsu_GN} }
    \label{fig:modulation}
\end{figure}

We use our library in three different scenarios: simple, with multiple routes, and considering modulation formats and optical reach. The first scenario is the simplest possible. There is no reach so that the algorithm can make connections independently of the route length. Also, each pair of nodes can be communicated only by one route. The second scenario is made with three possible routes: if the allocation process fails with the first route, then there is a second attempt with the second route. Finally, if the second route fails, there is a third attempt with the third route. The last scenario considers three routes, optical reach, and modulation format. The algorithm starts with the modulation format that uses fewer slots (more complex modulation format) and sequentially moves to the simplest modulation format, using more slots. That is made for each route.  

Finally, the optical reach and modulation formats are presented in Table~\ref{tabla:bitrate_modulation_fsu_GN}~\cite{calderon2020}.

\begin{table}[htbp]
    \begin{center}
        \begin{tabular}{|c|c|c|c|c|c|c|}
            \hline
            \multirow{2}{*}{\textbf{Modulation}}& \multirow{2}{*}{\textbf{optical reach [km]}}&\multicolumn{5}{ c| }{\textbf{Bitrates [Gbps]}} \\\cline{3-7} 
              && \textbf{10} & \textbf{40} & \textbf{100} & \textbf{400} & \textbf{1000} \\
            \hline \hline
            \textbf{BPSK}  &5520 & 1 & 4 & 8 & 32 & 80  \\ \hline
            \textbf{QPSK}  &2720 & 1 & 2 & 4 & 16 & 40\\ \hline
            \textbf{8-QAM} &1360  & 1 & 2 & 3 & 11 & 27 \\ \hline
            \textbf{16-QAM}&560 & 1 & 1 & 2 & 8  & 20 \\ \hline
            \textbf{32-QAM}&240 & 1 & 1 & 2 & 7  & 16 \\ \hline
            \textbf{64-QAM}&80 & 1 & 1 & 2 & 6  & 14 \\ \hline
        \end{tabular}
    \caption{Optical reach per modulation format and FSU requirements per bit-rate and modulation format pair, for a BER value equal to $10^{-6}$.}
    \label{tabla:bitrate_modulation_fsu_GN}
    \end{center}
\end{table}

We simulate $10^7$ connection requests, following a Poisson traffic with $\lambda \in \{18, 36, 54, 72, 90, 108, 126, 144, 162, 180\}$, and $\mu$ fixed in $10$. The bitrate requests follows a uniform distribution, where each bitrate is demanded with a probability equals to $\frac{1}{6}$.

The results show that the behavior of these algorithms is quite similar for the conditions tested within each scenario. It can be seen that the blocking probability curves decrease with increasing the number of routes used, and also with increasing the number of modulation formats because of the fewer number of slots required. The results got by the Flex Net Sim library in the different scenarios are as expected.

\section{Conclusion}

In this work, an event-oriented simulation library in flexible grid optical networks was presented. The library aims to facilitate the implementation and testing of new resource allocation algorithms in programs written in C ++. 

We implement three algorithms , which were executed in three different scenarios. The results show that the behavior of the library was as expected, reducing the probability of blocking in the scenarios that provided more opportunities for resource allocation. 

In the future, we expect to extend this library to implement algorithms in flexible grid optical networks in the domain of Spatial Division Multiplexing and Band Division Multiplexing.

\bibliographystyle{unsrt}  
\bibliography{references}  

\end{document}